\documentclass[preprint,10pt]{elsarticle}

\usepackage{amssymb}
\usepackage{upgreek}
\usepackage{color}
\usepackage{geometry} 
\usepackage[latin9]{inputenc}
\usepackage{color}
\usepackage[english]{babel}
\usepackage{calc}
\usepackage{graphicx}
\usepackage{amsmath}
\usepackage{textcomp}
\usepackage{hyperref}
\journal{ Diamond and Related Materials }

\begin{document}

\begin{frontmatter}
\ead{taras@physics.uq.edu.au}
\title{NV-centers in Nanodiamonds: How good they are \footnote{\copyright 2017 This manuscript version is made available under the CC-BY-NC-ND 4.0 license \url{http://criativecommons.org/licenses/by-nc-nd/4.0/}} }
\author{Taras Plakhotnik and Haroon Aman}
\address{School of Mathematics and Physics, The University of Queensland, St Lucia, QLD 4072, Australia}
\begin{abstract}
This paper presents a method for determination of the size distribution for diamond nanocrystals containing luminescent nitrogen-vacancy (NV) centers using the luminescence intensity only. We also revise the basic photo physical properties of NV centers and conclude  that the luminescence quantum yield of such centers is significantly smaller than the frequently stated 100\%.  The yield can be as low as 5\% for centers embedded in nanocrystals and depends on their shape and the refractive index of the surrounding medium. The paper also addresses the value of the absorption cross-section of NV centers.    
\end{abstract}


\end{frontmatter}

\section{Introduction}
Recent 20 years have seen explosive increase of publications on nitrogen-vacancy (NV) centers in diamond \cite{NV_rev}.  A simple search of \emph{Web of Science} reveals that the number of citations on the topic "nitrogen-vacancy" doubles approximately every 3 years and reached ten thousand in 2016 alone. The remarkable popularity of NV centers is based on their extraordinary photo-physical properties such as high photo-stability, sensitivity to magnetic fields, temperature and pressure, optical polarisation of electronic spin, optically detected magnetic resonance (ODMR) demonstrated even on  a single center at room temperature. 

Some of the applications of NV centers can profit greatly when the centers are imbedded in nearly perfect isotopically homogeneous crystals but other  are feasible only with nanodiamond. An obvious example of the later is labelling for bio imaging. A less obvious example is temperature measurements on a nanoscale.  Application of NV centers for temperature measurements in nano-structures is problematic with  bulk  crystals because very high thermal conductivity of diamond will inevitably affect the temperature at the location of interest. Nanocrystals of diamond photo activated with NV centers are commercially available from several suppliers. However current methods of production by irradiation, annealing and crashing of high-temperature high-pressure (HTHP) micro-sized crystals  result in nanocrystals which are much more inhomogeneous than other common bio labelling agents such as dye molecules or quantum dots. We will investigate  this inhomogeneity in the next section.    

A number of reviews are available on sensory applications and photo-physical properties of NV-centers \cite{NV_rev,  rev_1, rev_2, taras_NV_rev}  but readers should bear in mind that these reviews quickly age except for the parts dealing with most fundamental characteristics. In the following sections we will focus on some of these fundamentals to see if all the common claims about the NV-centers are sufficiently justified. 
 
\section{Brightness of nanodiamonds}
\begin{figure}[htbp] 
   \centering
   \includegraphics[width=8cm]{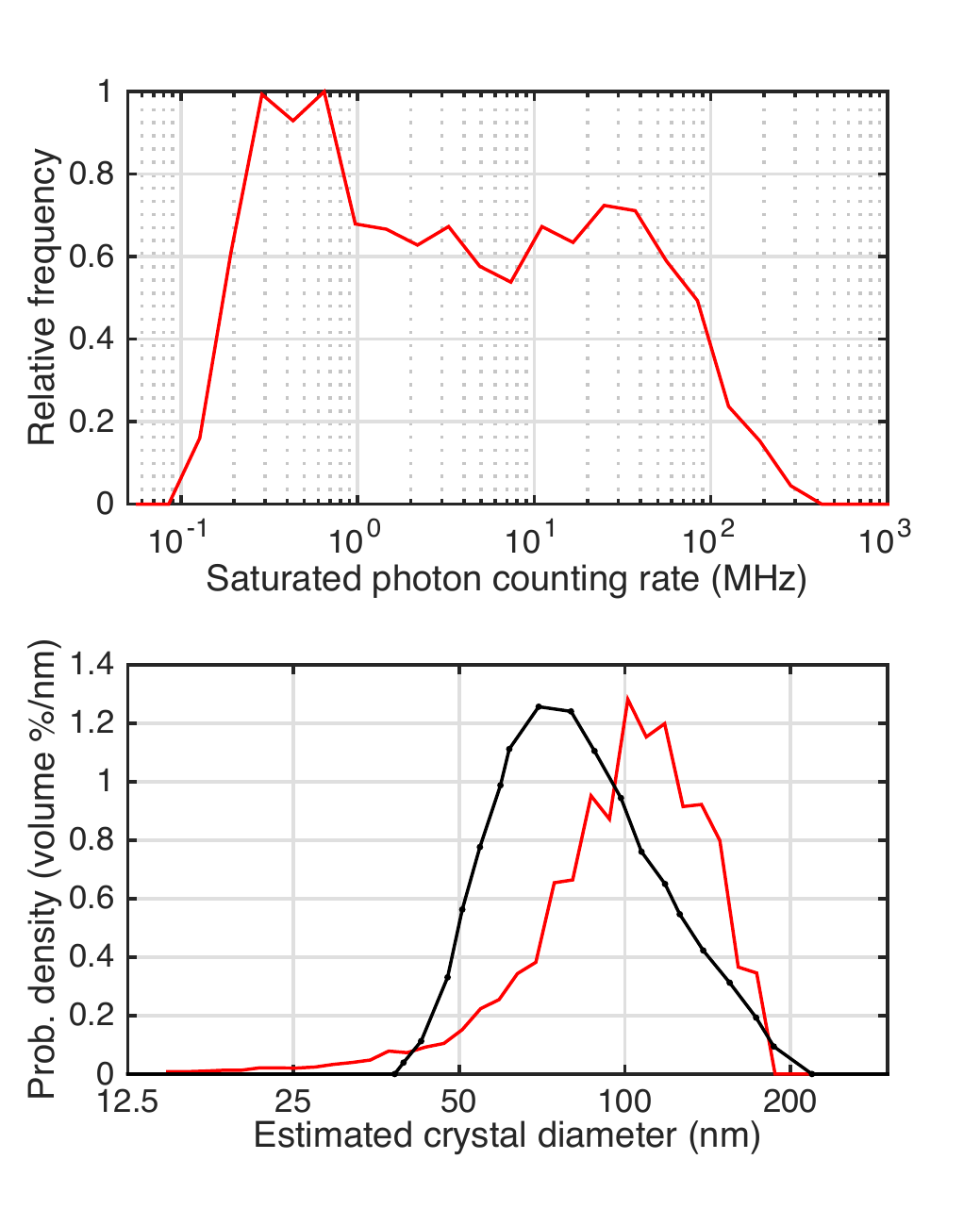} 
   \caption{Distribution of the count rates (top panel) and diameters of the crystals (bottom panel) in a drop of diamond suspension. The bottom panel shows DLS data (filled circles) for comparison. }
   \label{fig:distribution}
\end{figure}

Specific brightness (the maximum number of photons which can be detected from a unit volume of the material) is an important figure of merit for many applications.  This characteristic of luminescent nanodianmonds has been investigated using commercially available 70-nm fluorescent nanodiamond from Sigma-Aldrich. Product \# 798169 is supplied  in deionised water at mass concentration of diamond at 1 mg/mL  and advertised concentration of NV centers  of more than 300 per crystal. For the purpose of this research, the concentration of  diamond in the suspension has been decreased to $C_\mathrm{D}\approx 5\times10^{-4} \text{ mg mL}^{-1}$ (2000 times dilution) and a small drop of it (volume $V_\mathrm{drop}\approx 8.4\times10^{-5} \text{ mL}$)  has been deposited on a pre-cleaned glass microscope slide. Additionally, a number of water droplets has been placed around  the drop of suspension  to reduce formation of a coffee ring \cite{cofee-ring, coffee_ring_2}. After drying, the crystals were distributed approximately homogeneously on the slide within an approximately round spot  of about 1.2 mm in diameter. The photon count rate for each crystal has been measured under continuous 50-mW excitation at the wavelength of 532 nm (measured at the output of the exciting laser, Coherent Verdi-V5). The luminescence has been collected with a microscope objective, Numerical Aperture  (NA) of 0.9 and detected by a photon counting CCD (Andor iXon). The maximum power density of the excitation light beam on the slide has been estimated at about 3.5 kW/cm$^{2}$. The dependence of the detected count rate, $k^\mathrm{(det)}_\mathrm{r}$ on the excitation power $P$ has been measured for 9 nanocrystals and  followed a typical saturation curve \cite{sing_mol_rev}
\begin{equation}\label{eq:P_sat}
k^\mathrm{(det)}_\mathrm{r}=\mathcal{R}^\mathrm{(det)}\frac{P}{P+P_\mathrm{s}}
\end{equation}
The average saturation power, $P_\mathrm{s}$ for this set has been estimated at the level of 0.9(3)\,W for the case of a crystal in the center of the illuminated area (the corresponding average saturation intensity is $70\pm15 \text{ kW/cm}^2$ in agreement with \cite{taras_2010} but the standard deviation in the sample was about 30\% of the mean). The value of  $\mathcal{R}^\mathrm{(det)}$, the maximum detectable photon count rate for each crystals can be estimated using the count rates measured at the excitation power of 50 mW and the value of $P_\mathrm{s}$.  We have also taken into account the variation of the laser light irradiance across the  spot illuminated with the laser. This variation has been determined using weak luminescence of substrate which is proportional to the irradiance.  The relatively large uncertainty in the saturation power was a little problem for the following as the observed distribution of the saturated count rates has varied from crystal to crystal by a factor of 1000 as can be seen in the top panel of Fig.\ref{fig:distribution}. If the number of NV-centers per crystal is (as advertised) larger than 300 for each crystals, then some of the crystals should have about a million of such centers (unless many of these centers are very dim in the crystals with low $\mathcal{R}^\mathrm{(det)}$). To clarify this problem, we will analyse the size distribution of the crystals using the luminescence data.   

The total volume of the diamond in the drop can be deduced from the mass concentration of the diamond, the volume of the drop and the density of the diamond. This value should be equal to the sum of $\mathcal{R}^\mathrm{(det)}_j/ {\beta_j}$ over all the crystals in the drop      
\begin{equation}\label{eq:volume}
\frac{C_\mathrm{D}V_\mathrm{drop}}{\rho}=\sum_{j=1}^N\frac{\mathcal{R}^\mathrm{(det)}_j}{\beta_j},
\end{equation}
 where $\rho\approx 3.5\text{ g cm}^{-3}$ is the density of diamond crystal and $\beta_j$ is the the maximum number of photons detected from $j$-th crystal divided by its volume. Assuming that $\beta_j\equiv \beta$ is the same for all the crystals, one can find the value of $\beta$  from Eq. (\ref{eq:volume}) and then determine a diameter of each detected crystal using their volumes $V_j\approx d_j^3\pi/6$ and the corresponding maximum detectable rates $\mathcal{R}_j^\mathrm{(det)}$. 

\begin{equation}
d_j=\left(\frac{6C_\mathrm{D}V_\mathrm{drop}\mathcal{R}_j^\mathrm{(det)}}{\pi\rho\sum_{j=1}^N\mathcal{R}_j^\mathrm{(det)}}\right)^{1/3}
\end{equation}
 The result of such calculations is shown in the bottom panel of Fig.\ref{fig:distribution}. The distribution of the diameters measured by dynamic light scattering (DLS) is also shown and demonstrates  quite a reasonable agreement with the distribution obtained from the luminescence measurements (especially given the assumptions and approximations made in the data analysis). The estimated value of $\beta$ is 150 Hz/nm$^3$. In addition to the inaccuracy of the approximations introduced above, a small mismatch between the position of the maxima in the two distribution (70 and 100 nm) can be explained by the assumption that some of  the crystals do not have detectable NV-centers (strictly speaking $C_\mathrm{D}$ should refer  to detectable diamonds only).  
 
 Given that about 10\% of emitted photons are detected by our setup (see more details below), the estimated value of   $\beta$ corresponds to an absolute specific brightness of about $1.5\times10^3$ photon/nm$^3$.  If we take the mean diameter of the crystals  (averaged by volume) to be 100 nm (in agreement with the distribution shown in Fig.\ref{fig:distribution}) and the advertised 300 NV centers as the value for such an average size, then the emission photon rate is 2.5 MHz per NV-center. 

In summary, this experiment shows that commercially available diamonds can be characterised by  a specific brightness. This observation suggests a uniform concentration of NV-centers as a reasonable approximation. The wide inhomogeneity of the overall brightness mainly results from a wide distribution  of the crystal volumes in the supplied material.  

But does the observed specific brightness  matches up with what we know about the photophysical properties of the NV-centers? Note that 2.5 MHz is 16 times smaller than a frequently reported luminescence decay rate of 40 MHz (25 ns lifetime). The answer to this question will be one of the main subject of the following discussion.   

\section{Interaction with electromagnetic fields }
The interaction of photons with an NV center embedded in a crystal of diamond presents a significant theoretical challenge at the level of microscopic fields inside the crystal.  But when these microscopic fields are averaged, one gets macroscopic fields  which are much easier to calculate. In principle, identical macroscopic fields can correspond to different microscopic fields but one can assume a one-to-one correspondence between macroscopic and microscopic fields acting on a single NV center in a perfect diamond crystal. If the center resides in a  sufficiently large nanocrystal and we can ignore interaction of the NV with any other impurities or defects then equal macroscopic fields in the nanocrystals and in  bulk  will correspond to equal microscopic fields acting on the NV-center. We will use this equality in the  following review of the basic principles of NV-photon interaction and the corresponding dynamics of populations. 

\subsection{Absorption cross-section}
The optical absorption cross-section is defined by the equality
\begin{equation}\label{eq:cross-section}
k_{lu}=\sigma I
\end{equation}
where $k_{lu}$ is the excitation rate from the lower electronic state to the upper electronic state  and $I$ the photon flux  (expressed in units of photon\,s$^{-1}$\,m$^{-2}$). Note that $k_{lu}$ multiplied by the energy difference between the upper and the lower states gives the absorbed power. One should keep in mind that the excitation rate for a typical case of an electric-dipole transition is proportional to $\cos^2\theta$,  where $\theta$ is the angle between the microscopic electric field and the direction of the transition dipole. The NV-centers have two mutually orthogonal electrical-dipole transitions due to two electronically excited degenerate states $^3\mathrm{E}$ (see Subsection \ref{sec:population} and Fig.\,\ref{fig:NV_model}). This creates additional complications for the analysis of absorption by a single center as neither the orientation of the dipoles nor the direction of the microscopic field may be easy to determine. For simplicity and if not explicitly stated otherwise, we will assume that  $\sigma$  is the absorption cross-section averaged under the assumption of uniform probably distribution for the relative orientation of the dipole axis and the direction of the microscopic field. This averaging is partially facilitated by the presence of the two orthogonal dipoles but the averaged quantity would rather refer to a property of a sample with many centers. When we  compare properties of an NV-center  embedded in  bulk  diamond and in nanocrystals, it  will be assumed that the corresponding angles between the dipoles and the electrical field  are equal. 

First, we consider interaction of NV center imbedded in a  bulk  diamond with a plane wave entering the diamond from air perpendicular to the interface. On the interface, the wave is partially reflected and $E_\mathrm{i}$, the macroscopic electrical field inside the diamond is reduced in comparison to the external macroscopic field $E_\mathrm{ext}$: 
\begin{equation}
E_\mathrm{i[b]}=\frac{2}{n+1}E_\mathrm{ext}=\eta_\mathrm{[b]}E_\mathrm{ext}
\end{equation}
where $n\approx 2.42$ is the refractive index of diamond. The factor $\eta$  will be called a shielding factor (in this case for a  bulk  crystal as indicated by the subscript).  Note that this is a much larger effect than the photon flux  loss due to the reflection. The transmitted power is just $4n/(n+1)^2$ times the incident power. The microscopic field  interacting with  the NV center  is proportional to $E_\mathrm{i[b]}$ and therefore the absorbed power is proportional to $E_\mathrm{i[b]}^2$. 

Because calculation of  macroscopic fields inside of an irregularly shaped crystal can be done only numerically, we limit the analysis to  interaction of an NV center with an electro-magnetic wave incident on an ellipsoid-shaped crystal with arbitrary relation between its three axes. The lengths of these axes are $a$, $b$ and $c$, with $a$ being the longest. The electrical field inside an ellipsoid is homogeneous and  this makes the analysis particularly simple. The assumption that the polarisation of the wave is directed along one of the ellipsoid axes, does not limit the generality of the theory due to the validity of the linear superposition principle  for the low field strength typical for this situation.  When the external field is directed along one of the axes, the internal field is parallel to the external field as in the case of a plane interface but the proportionality constant is different.   
\begin{equation}\label{eq:ellipsoid_field}
E_\mathrm{i[el]}=\frac{1}{1+(n^2-1)\delta_\mathrm{[el]}}E_\mathrm{ext}=\eta_\mathrm{[el]} E_\mathrm{ext}
\end {equation}
where  the value of the depolarising factor $\delta$ depends on the shape of the ellipsoid and the direction of the field \cite{depolarisation}. For a sphere, $\delta=1/3$. The value of  $\delta$  is shown  in Tab.\ref{tab:ellipsoid} for a few characteristic cases. Numerical simulations show less than 10\% difference between a cylindrical disk and  a similarly sized  ellipsoid where one of the axis equals the length of the cylinder and the other two its diameter.

The rate of energy absorption by an NV center is proportional to the square of the microscopic field interacting with the center.  Because the macroscopic fields in  bulk  and inside  the ellipsoid are parallel to each other, the corresponding microscopic fields are also parallel and their magnitudes are  proportional to  the magnitudes of the macroscopic fields. Hence (see Eq.(\ref{eq:cross-section}))
\begin{equation}
\frac{k_\mathrm{lu[el]}}{k_\mathrm{lu[b]}}=\frac{|E_\mathrm{i[el]}|^2}{|E_\mathrm{i[b]}|^2} =\frac{\sigma_\mathrm{[el]}I_\mathrm{ext}}{\sigma_\mathrm{[b]}I_\mathrm{i[b]}}
\end{equation}
We use the internal flux of photons in the case of  bulk  (it is a usual practice to take into account the reflection on the interface between diamond and a surrounding medium and use $E_\mathrm{i[b]}$ in the calculations of the cross-section) but internal flux  is a meaningless concept for a nanocrystal. The absorption cross-sections in a  bulk  sample and in a nanocrystal are related as follows  
\begin{equation}\label{eq:sigma}
\sigma_\mathrm{[el]}=\sigma_\mathrm{[b]}\left(\frac{\eta_\mathrm{[el]}}{\eta_\mathrm{[b]}}\right)^2\frac{4n}{(n+1)^2}=\sigma_\mathrm{[b]}\eta^2_\mathrm{[el]}n
\end{equation}
The factor $I_\mathrm{i[b]}/I_\mathrm{ext}=4n/(n+1)^2$ is the energy flux transmittance  of the   bulk  diamond-air interface at a normal incident (numerically this factor is about 0.83 for a diamond-air interface). Equation (\ref{eq:sigma}) can  also be applied to analyse a practically important case of nanodiamonds dispersed in water but the refractive index of diamond $n$ should be replaced with $n/n_\mathrm{w}$, where $n_\mathrm{w}$ is the refractive index water. 

Values of $\eta^2_\mathrm{[el]}n$ for directions of the electric field along the ellipsoid axes are shown in Tab.\,\ref{tab:ellipsoid} for diamond surrounded by air or water. The factor $\delta$ depends only on the shape of the ellipsoid. The table also includes these values averaged over the field directions (this approximately simulates absorption of an ensemble of randomly oriented nanocrystals). 

One has to be careful when applying Eq.(\ref{eq:sigma}) to a nanocrystal siting on a substrate (a glass slide, for example). In such a case, the external field is significantly affected by the reflection of the incoming wave from the surface of the substrate.  Interference between the incoming and reflected waves reduces the electrical field. The field inside large but very thin flake of diamond (much thinner than the wavelength inside the diamond)  on a glass surface will be equal to $\eta_\mathrm{[s]}E_\mathrm{[ext]}$ (as follows from the boundary conditions). Analytical solution of the corresponding electrostatic problem   is available for a sphere (a valid approximation if the sphere is much  smaller than the wavelength of light) but is quite complicated \cite{sphere_plane_elertstatics}. We will discuss numerical simulations later in this paper. 

\subsection{Photon emission rate}
In this subsection we consider spontaneous radiative transitions and compare radiative rates of  NV centers imbedded in  bulk  diamond and in a nanocrystal.  The density of the photon states in a homogeneous not absorbing and not dispersive media is proportional to $n^3$ while the strength of the macroscopic field associated with one phonon is proportional to $n^{-1}$ \cite{photon_medium}.  Suppose that  a crystal is too small to support any localised optical modes in the relevant spectral range.The photon modes which interact with the NV-center will be then basically the photons of the vacuum whose density of states is $n^3$ times smaller than the density of states in the  bulk  crystal but the associated electrical field is $n$ times stronger. This macroscopic field will be reduced inside the nanocrystals  and will be equal to $E_\mathrm{i[el]}$ as defined by Eq.(\ref{eq:ellipsoid_field}) where the value of $E_\mathrm{ext}$ is the field associated with a single photon mode in vacuum.  In the case of a very small spherical  particle,  the rate of the spontaneous emission $k_\mathrm{r[s]}$ and the rate of the radiative decay of  NV center in  bulk  diamond $k_\mathrm{r [b]}$ are related as follows \cite{Chew_1988}
\begin{equation}\label{eq:rad_sphere}
k_\mathrm{r[s]}=\frac{\eta^2_\mathrm{[s]}}{n} k_\mathrm{r[b]}= \left(\frac{3}{2+n^2}\right)^2\frac{1}{n}k_\mathrm{r[b]}
\end{equation}
where $\eta_\mathrm{[s]}$ is the value of $\eta$ calculated for a sphere. The validity of Eq.(\ref{eq:rad_sphere}) has been demonstrated experimentally \cite{Sandoghdar_2002}.  To calculate the radiative rate of a dipole transition (the unit vector of the dipole is $\hat{\mu}$) in the case of an elliptical particle, one has to average $\eta^2 \cos^2\theta$. The angle $\theta$ is the angle between   the direction of the electric field and $\hat{\mu}$.  To compute the average, one has to consider three orthogonal directions of the wavevector, chosen along  the axes of the ellipsoid.  In a simple case of $\hat{\mu}$ parallel to axis $a$,  $\cos\theta= 1$  only when the electric field of the wave is also parallel to $a$. The wavevector of such a wave should be therefore parallel to the axes $b$ or $c$. For other waves/polarisations, $\cos\theta= 0$. The electric-dipole decay rate in a nanocrystal reads 
\begin{equation}\label{eq:k_rad}
k_\mathrm{r[el]}= \frac{\left\langle\eta_\mathrm{[el]}^2\right\rangle}{n} k_\mathrm{r[b]}
\end{equation} 
The values of $\left\langle\eta_\mathrm{[el]}^2\right\rangle/n$  for three directions of the transition dipole moment (along $a$, $b$, and $c$ axis) are shown in Tab.\ref{tab:ellipsoid} for air or water surroundings.  In the case of water, $n$ is the refractive index of diamond relative to water. The average value gives an indication of the decay rate in the case of many centers randomly oriented within ellipsoids (but strictly speaking the decay is not single exponential in this case). Even if we exclude the extreme cases such as a needle ($b,c=0$) or a flake ($c=0$) from  Tab.\,\ref{tab:ellipsoid}, the radiative rates in   bulk   diamond and in a nanocrystal (surrounded by air) differ by a factor between 0.016 and 0.22. The actual  shapes of the  crystals are quite irregular but to narrow the range a bit we will use $\left\langle\eta_\mathrm{[el]}^2\right\rangle/n$  limited to the cases when the shortest axis is not shorter than 1/2.  For such a case, $k_\mathrm{r[nano]} \approx 0.07(3)\times k_\mathrm{r[b]}$ and $k_\mathrm{r[nano]} \approx 0.19(5)\times k_\mathrm{r[b]}$ are estimates for the radiative rates if the crystals are surrounded by air and water respectively. The mean and the standard deviation (in brackets) are obtained using the 9 values of $\eta_\mathrm{[el]}^2/n$ in the table (see sections $b,c=1/2$, $c=1$, and $c=1/2$).

In the case of the nanocrystal placed on a dielectric substrate (refractive index $n_\mathrm{sub}$), the proximity of the interface  will affect the radiative rate. The magnitude of the effect depends on the orientation of the electronic transition dipole of the NV center and its distance from the interface. If the transition dipole is parallel to the plane, the effect is relatively small even when the dipole sits exactly on the interface \cite{dipole_em_nearplane, dipole_surf_Mertz}. In such a case  the increase in the emission rate is  about a factor of 1.4 for a typical value of $n_\mathrm{sub}\approx 1.5$. The rate increases by a factor of about 2.3 if the transition dipole is  perpendicular to the plane. Note that the excited triplet state of  NV centers is degenerated and the two corresponding transitions have mutually orthogonal transition dipoles. This makes the effect of the interface less dependent on the orientation of the center. Numerical calculations require a number assumptions but an average value  of 1.7  is a reasonable estimate  for a crystals on a glass substrate with a large number of NV-centers \cite{spont_emission_surf}. This is the averaged effect for the  three mutually orthogonal orientations of the transition dipoles (two in the plane and one perpendicular to the interface). 

 \begin {table}
\begin{tabular}{|c |l | c | c | c | c | l}
  \hline
  Shape 			&				&$E\parallel a$&$E\parallel b$	&$E\parallel c$& av.\\
 \hline
  	   			&$\delta_\mathrm{[el]}$		& 0 		& 0.50 	& 0.50 	&--\\
   $b,c=0$   		&$\eta_\mathrm{[el]}^2 n$		&2.4		&0.21	&0.21	&0.94\\
	$(a=1)$		&--						&1.8		&0.39	&0.39	&0.87\\
				&$\eta_\mathrm{[el]}^2/n$		&0.41	&0.035	&0.035	&0.16\\
				&--						&0.55	&0.12	&0.12	&0.26\\
\hline
  				&$\delta_\mathrm{[el]}$		& 0.075	& 0.46	&0.46	&--\\
  $b,c=\frac{1}{4}$	&$\eta_\mathrm{[el]}^2 n$		&1.3		&0.23	&0.23	&0.59\\
  	$(a=1)$		&--						&1.3		&0.42	&0.42	&0.72\\
 				&$\eta_\mathrm{el}^2/n$		&0.22	&0.039	&0.039	&0.10\\
				&--						&0.40	&0.13	&0.13	&0.22\\
				\hline
  				&$\delta_\mathrm{[el]}$		&0.17	& 0.41	&0.41	&--\\
 $b,c=\frac{1}{2}$ 	&$\eta_\mathrm{[el]}^2 n$		&0.71	&0.27	&0.27	&0.42\\
 	$(a=1)$		&--						&0.93	&0.48	&0.48	&0.62\\
 				&$\eta_\mathrm{[el]}^2/n$		&0.12	& 0.046	&0.046	&0.071\\
				&--						&0.28	&0.14	&0.14	&0.19\\
\hline
  	    			&$\delta_\mathrm{[el]}$		&0.33	&0.33	&0.33	&--\\
  $b,c=1$		    	&$\eta_\mathrm{[el]}^2 n$		&0.35	&0.35	&0.35	&0.35\\
  $(a=1)$			&--						&0.58	&0.58	&0.58	&0.58\\
				&$\eta_\mathrm{[el]}^2/n$		&0.060	&0.060	&0.060	&0.060\\
				&--						&0.18	&0.18	&0.18	&0.18\\

\hline
  				&$\delta_\mathrm{[el]}$		&0.24	&0.24	&0.53	&--\\
  $c=\frac{1}{2}$	&$\eta_\mathrm{[el]}^2 n$		&0.52	&0.52	&0.19	&0.41\\
				&--						&0.76	&0.76	&0.37	&0.63\\
$(a=b=1)$			&$\eta_\mathrm{[el]}^2/n$		&0.090	&0.090	&0.033	&0.071\\
				&--						&0.23	&0.23	&0.11	&0.19\\	
\hline
  				&$\delta_\mathrm{[el]}$		&0.15	&0.15	&0.70	&--\\
 $c=\frac{1}{4}$ 	&$\eta_\mathrm{[el]}^2 n$		&0.82	&0.82	&0.12	&0.59\\
$(a=b=1)$			&--						&1.0		&1.0		&0.26	&0.76\\
				&$\eta_\mathrm{[el]}^2/n$		&0.14	&0.14	&0.021	&0.10\\
				&--						&0.31	&0.31	&0.080	&0.23\\	
				\hline
 				&$\delta_\mathrm{[el]}$		&0.085	&0.085	&0.70	&--\\
   $c=\frac{1}{8}$	&$\eta_\mathrm{[el]}^2 n$		&1.22	&1.22	&0.096	&0.84\\
   $(a=b=1)$		&--						&1.27	&1.27	&0.21	&0.92\\
				&$\eta_\mathrm{[el]}^2/n$		&0.21	&0.21	&0.016	&0.14\\
				&--						&0.39	&0.39	&0.065	&0.28\\	
\hline
  				&$\delta_\mathrm{[el]}$		&0		&0		&1.0		&--\\
  $c=0$			&$\eta_\mathrm{[el]}^2 n$		&2.4		&2.4		&0.071	&1.6\\
  $(a=b=1)$			&--						&1.8		&1.8		&0.17	&1.3\\
				&$\eta_\mathrm{[el]}^2/n$		&0.41	&0.41	&0.012	&0.28\\
				&--						&0.55	&0.55	&0.050	&0.38\\
 \hline  
\end{tabular}
\caption {Depolarisation, cross-section correction, and radiative decay factors for ellipsoids. The factors are calculated for each shape (defined by the relative lengths of axes $a$, $b$, and $c$) and three directions of the filed. Row (1) shows values of $\delta$;  (2) and (3) show the effect of the ellipsoids on the absorption cross-section for the case of water and air surroundings respectively;  Rows (4,5)  display the effect of the ellipsoids on the radiative rates (water, air).}  \label{tab:ellipsoid} 
\end{table}

A widely used relation between the absorption cross-section and the radiative lifetime   \cite{rate_to_sigma} reads 
\begin{equation}\label{eq:rad_to_sigma}
k_\mathrm{r}=n^2 \frac{8 \pi c }{ \langle \tilde{\nu}^{-3}_\mathrm{lum} \rangle}\frac{g_l}{g_u}\sigma_\mathrm{max}\int\frac{\bar\sigma(\tilde{\nu}_\mathrm{abs}) }{\tilde{\nu}_\mathrm{abs}} d\tilde{\nu}_\mathrm{abs}
\end{equation}
where $\tilde{\nu}_\mathrm{abs}$  and $\tilde{\nu}_\mathrm{lum}$  are absorption and luminescence frequencies in units of wavenumbers (cm$^{-1}$),  $ \langle \tilde{\nu}_\mathrm{lum}^{-3} \rangle \equiv \int s(\tilde{\nu}_\mathrm{lum})\tilde{\nu}_\mathrm{lum}^{-3} d\tilde{\nu}_\mathrm{lum}/\int s(\tilde{\nu}_\mathrm{lum}) d\tilde{\nu}_\mathrm{lum}$ is the value of $ \tilde{\nu}_\mathrm{lum}^{-3}$ averaged using luminescence spectral density  $ s(\tilde{\nu}_\mathrm{lum})$ as a weighting factor. The absorption spectrum $\bar\sigma$ is normalised so that it equals 1 at the maximum and  is  averaged over all equally probable orientations of the transition dipole moment (this may be not strictly valid for a crystal fixed in the laboratory reference frame  when the absorption cross-section is measured for a particular direction of the incident light). This equation is derived by considering a thermodynamic equlibrium  between blackbody radiation and a quantum system. The refractive index $n$ is the refractive index of the media which is filled with black body radiation. In the case of a  bulk  diamond $n=2.42$  while $n=1$ for the nanocrystals in vacuum (air).  The derivation also relies on the validity of the Frank-Condon principle for transition probabilities between quantum vibronic  states. The expression does not include the local field factor because we have assumed the same factor for absorption and emission fields. This  strictly speaking is not an exact statement (see, for example, \cite{Stoneham_1975} for discussion). The challenging problem of local fields does not arise in relation to Eq.(\ref{eq:sigma}) and Eq.(\ref{eq:k_rad}) as they both deal with only absorption or only emission (only the size and/or shape of the crystal change). For a crystal on a substrate, the radiative rate is higher but Eq.(\ref{eq:rad_to_sigma}) is not valid as the refractive index is not homogeneous.  One can use  Eq.(\ref{eq:sigma}) and Eq.(\ref{eq:k_rad}) to derive that $k_\mathrm{r[el]}/\sigma_\mathrm{[el]}=n^2 k_\mathrm{r[b]}/\sigma_\mathrm{[b]}$. This equality also follows from Eq.(\ref{eq:rad_to_sigma}) and thus it proves the consistency between electrodynamics and thermodynamics considerations.  

\subsection{Population dynamics}\label{sec:population}
The structure of quantum states in NV-centers is quite complicated. A minimally realistic approach includes 3 spin sub-levels of the  ground electronic orbital, $3\times2$  sub-levels of the double degenerated electronic orbitals of the exited triplet and a couple of singlet orbitals. In a more accurate model, one also should consider the vibrationally excited states to account for the absorption of 532-nm wavelength  of the exciting light and states which can be reach by two-photon excitation. All this makes the complete analysis very involved even in the approximation of the rate equations. However, important relations can be derived by making a few approximations. 
\begin{figure}[htbp] 
   \centering
   \includegraphics[width=5cm]{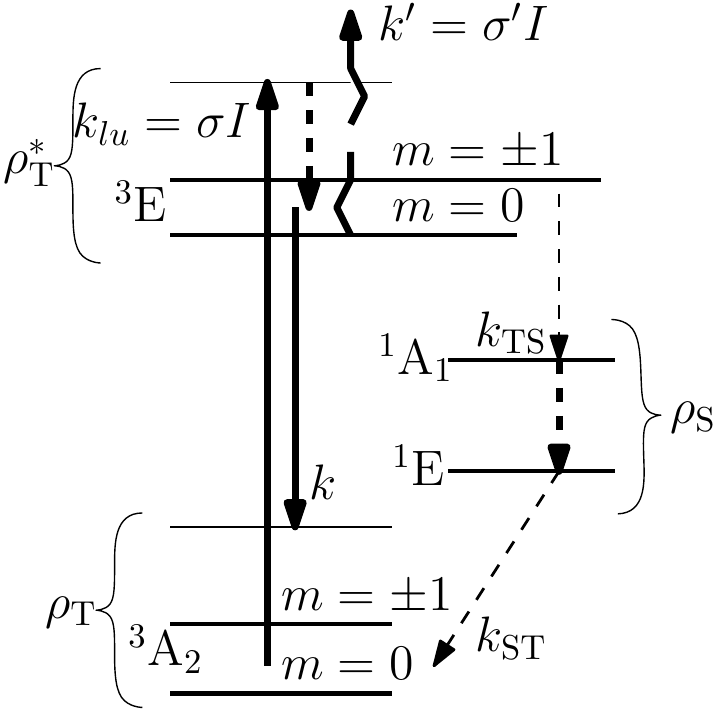} 
   \caption{A reduced model of an NV center. The rates shown by dashed lines are not radiative. The fat dashed lines are considered so fast that the populations of their upper states are negligible. The relative population of $m=\pm1$ and $m=0$ spin sub-levels is characterised by the parameter $\alpha$. The figure indicates that the intersystem crossing takes place from the $m=\pm1$ sub-levels of the $^3\mathrm{E}$ degenerate electronic states. The rate $k$ the sum of  both radiative $k_\mathrm{r}$ and not radiative $k_\mathrm{nr}$  rates. The symmetry of the electronic orbitals and their spin values are also indicated.}
   \label{fig:NV_model}
\end{figure}

First,  the $^1\mathrm{A}_1$ singlet state lifetime is very short ($\approx 1$ ns) in comparison to the $^1\mathrm{E}$ singlet (about 300 ns at room temperature). Therefore we can neglect population of $^1\mathrm{A}_1$ and write a simple rate balance between the populations of the excited triplet state  $\rho_\mathrm{T}^*$ and the lowest singlet $\rho_\mathrm{S}$ as follows  
\begin{equation}\label{eq:tr_pop}
\alpha\rho_\mathrm{T}^*k_\mathrm{TS}=\rho_\mathrm{S}k_\mathrm{ST}
\end{equation}
where   the phenomenological parameter $\alpha$ characterises the relative population of the $m=\pm1$ sublevels of the $^3\mathrm{E}$ orbital (we also assume that the intersystem crossing is dominated by the transitions from the $m=\pm1$ levels). In this equation, the left hand side is the rate of singlet population due to intersystem crossing from the excited triplet and  the right hand side is the rate of depopulation of the singlet. The population of the singlet state can be obtained from Eq.(\ref{eq:tr_pop}) as follows $\rho_\mathrm{S} = \alpha\rho_\mathrm{T}^* k_\mathrm{TS}/k_\mathrm{ST}$. 

Second, we note that a typical relaxation time from the vibrationally excited states down is  several picoseconds, orders of magnitude shorter than the relaxation time from $^3\mathrm{E}$ to other electronic orbitals (about 10 ns). Therefore we can formally consider direct excitation to the electronic  triplet $^3\mathrm{E}$. The complication of the orbital  degeneracy is simplified at room temperature by the very fast  phonon mitigated exchange of the populations between the two orbitals. The rate of optical excitation to $^3\mathrm{E}$ is $k_\mathrm{lu}=I \sigma \rho_\mathrm{T}$, where  $I$ is the intensity of the exciting light, $\sigma$ is the absorption cross-section, and $ \rho_\mathrm{T}$ is the population of the ground triplet. In a similar fashion, we consider excitation from the excited triplet to a higher electronic orbital followed by a fast relaxation to the ground state (we include this possibility for a reason which will become clear later). The steady state population of the excited triplet is reached when the balance between the depopulation and repopulation of the  ground triplet state holds, that is when $I\sigma\rho_\mathrm{T}=\rho_\mathrm{T}^*(\alpha k_\mathrm{TS}+k+I\sigma')$. Assuming that  $\rho_\mathrm{S}+\rho_\mathrm{T}^*+\rho_\mathrm{T}=1$ and $\rho_\mathrm{S} = \alpha\rho_\mathrm{T}^* k_\mathrm{TS}/k_\mathrm{ST}$, one can  express the value of  $\rho_\mathrm{T}^*$ as follows
\begin{equation}\label{eq:ex_tripl_pop}
\rho_\mathrm{T}^*=\frac{\sigma I}{\sigma I \left(\frac{\sigma+\sigma'}{\sigma}+\frac{\alpha k_\mathrm{TS}}{k_\mathrm{ST}}\right)+k+\alpha k_\mathrm{TS}}
\end{equation}
The detected photon rate equals $k_\mathrm{r}\Phi_\mathrm{det}\rho_\mathrm{T}^*$,  where $k_\mathrm{r}$  is the radiative decay rate  and $\Phi_\mathrm{det}$ is the efficiency of photon detection which is a product of the photodetector efficiency $\Phi_\mathrm{pd}$ and the  efficiency of the photon-collecting optics $\Phi_\mathrm{opt}$. 
The value of $\Phi_\mathrm{opt}$ can be calculated in many important cases analytically or numerically.  The collection efficiency  for a single dipole in homogeneous media can be found in \cite{Taras_1995} and for a dipole on a dielectric interface in \cite{pulsed_excit}. As a benchmark for this value we have calculated the efficiency of a standard microscope objective (NA=0.9, transmission 0.8, used in the experiments reported in this paper)  averaged over all orientations of NV centers in a  bulk  crystal and in a nano-diamond on a glass substrate.  In these two cases,  $\Phi_\mathrm{opt}=0.023$    and $\Phi_\mathrm{opt}=0.14$ respectively. Note that the much smaller value of  $\Phi_\mathrm{opt}$ for  bulk  diamond is mainly due to the decreasing of the numerical aperture by a factor of $n$ (the refraction index of diamond) due to refraction on the interface between the diamond and the air. The reflection on the diamond surface reduces the efficiency only by a factor of 0.8.

The maximum detected photon rate then reads (see Eq. (\ref{eq:P_sat})) 
\begin{equation}\label{eq:rate}
\mathcal{R}^\mathrm{(det)}=k_\mathrm{r}\Phi_\mathrm{det}\frac{k_\mathrm{ST}}{\frac{\sigma+\sigma'}{\sigma}k_\mathrm{ST}+\alpha k_\mathrm{TS}}
\end{equation}

Another important quantity is the so-called saturation intensity. This is the excitation intensity when the photon rate equals one half of $\mathcal{R}^\mathrm{(det)}$. The saturation intensity in units of W/cm$^2$ can be obtained from Eq.(\ref{eq:ex_tripl_pop}) as 
\begin{equation}\label{eq:sat}
I_\mathrm{s}=\frac{hck_\mathrm{ST}}{\sigma\lambda}\frac{\alpha k_\mathrm{TS}+k}{\alpha k_\mathrm{TS}+\frac{\sigma+\sigma'}{\sigma}k_\mathrm{ST}}
\end{equation}
where $hc/\lambda$ is the energy of one photon at the excitation wavelength $\lambda$. The saturation intensity monotonically increases with decreasing value of $\alpha$ (note that $k\ll k_\mathrm{ST}$).  

A combination of  $\mathcal{R}^\mathrm{(det)}$ and the saturation intensity gives a way to link the absorption cross-section to the quantum yield of NV luminescence. 
\begin{equation}\label{eq:sigma_qy}
\frac{\lambda}{hc}\frac{I_\mathrm{s}\Phi_\mathrm{det}}{\mathcal{R}^\mathrm{(det)}}=\frac{1}{\sigma \Phi_\mathrm{NV}}\times\left(1+\alpha \frac{k_\mathrm{TS}}{k}\right)\approx \frac{1}{\sigma \Phi_\mathrm{NV}}
\end{equation}
where $\Phi_\mathrm{NV}\equiv k_\mathrm{r}/k$ is the luminescence yield of the NV center. Because the value of $k_\mathrm{TS}/k \approx 0.5$ in  bulk  (it is about 1 in nano crystals), the value of the  term in the brackets  varies  only between 1 and 1.3 if $\alpha$ changes from 0 (the minimal possible value)  to 2/3 (its thermal equilibrium value). As a matter of fact, one can reduce the uncertainty in the value of $\alpha$ by using the ODMR data. We estimate $C_\mathrm{ODMR}$, the contrast of the ODMR line in a case of a low excitation intensity, such that the population of the ground triplet state is close to 1. 

To obtain an expression for $C_\mathrm{ODMR}$, we assume that the conditional probabilities of populating the  $m=\pm1$ and $m=0$ sub-levels are $\alpha$ and $1-\alpha$ respectively even at  the low excitation intensity, typical for the ODMR measurements.  In this approximation, every excitation to $^3\mathrm{E}$ results in emission of a photon with probability of 
\begin{equation}\label{eq:photon_prob}
p=(1-\alpha) \frac{k_\mathrm{r}}{k}+\alpha\frac{k_\mathrm{r}}{k+k_\mathrm{TS}}=\frac{(1-\alpha) k_\mathrm{r}k_\mathrm{TS}+k_\mathrm{r}k}{k(k+k_\mathrm{TS})}
\end{equation}
When NV-center is subject to strong  MW radiation at resonance with the $\langle0\vert \leftrightarrow \langle\pm1\vert$ transitions, the steady-state value of $\alpha$  equals 2/3  (this value implies that all three spin sub-levels are equally populated) and the corresponding probability  of the photon emission (as given by Eq. (\ref{eq:photon_prob})) is equal to $p_0$. The expression for the contrast then reads
\begin{equation}
C_\mathrm{ODMR} \equiv  \frac{p-p_0}{p}=\frac{ 2/3 - \alpha}{1+k/k_\mathrm{TS}-\alpha }
\end{equation}
Apparently, the contrast of the ODMR line can be used to estimate the value of $\alpha$.
 \begin{equation}\label{eq:alpha}
\alpha =\frac{2/3-C_\mathrm{ODMR}(1+k/k_\mathrm{TS})}{1-C_\mathrm{ODMR}}
\end{equation}

A typical value of $C_\mathrm{ODMR} \approx 0.25$  and $k/k_\mathrm{TS}\approx1$, observed for nanodiamonds, results in an estimate of $\alpha\approx 0.2$. Therefore the error due to approximation in Eq.(\ref{eq:sigma_qy}) is less than 20\%. 
\section{Discussion}\label{discussion}
The absolute value of the absorption cross-section of NV-centers is hard to estimate using conventional methods founded on Beer-Lambert law because the concentration of these centers can not be obtained easily. One of the most recent estimates for the absorption cross-section at 532-nm wavelength \cite{abs_bulk}  reports $\sigma_\mathrm{[b]}=(3.1\pm0.8)\times 10^{-17}\text{ cm}^2$. This estimate is based on the integrated ZPL absorption at 80 K temperature \cite{NV_abs_ESR_cal} in a sample where the concentration of the NV centers is measured using electron-spin resonance (ESR) data. 
The wavelength dependence of absorption and luminescence has been measured many times and is similar for nanocrystals and  bulk  material (if the micro-crystal resonates at particular frequencies, the shapes of the emission/absorption bands can be significantly modified). 

\begin{figure}[htbp] 
   \centering
   \includegraphics[width=8cm]{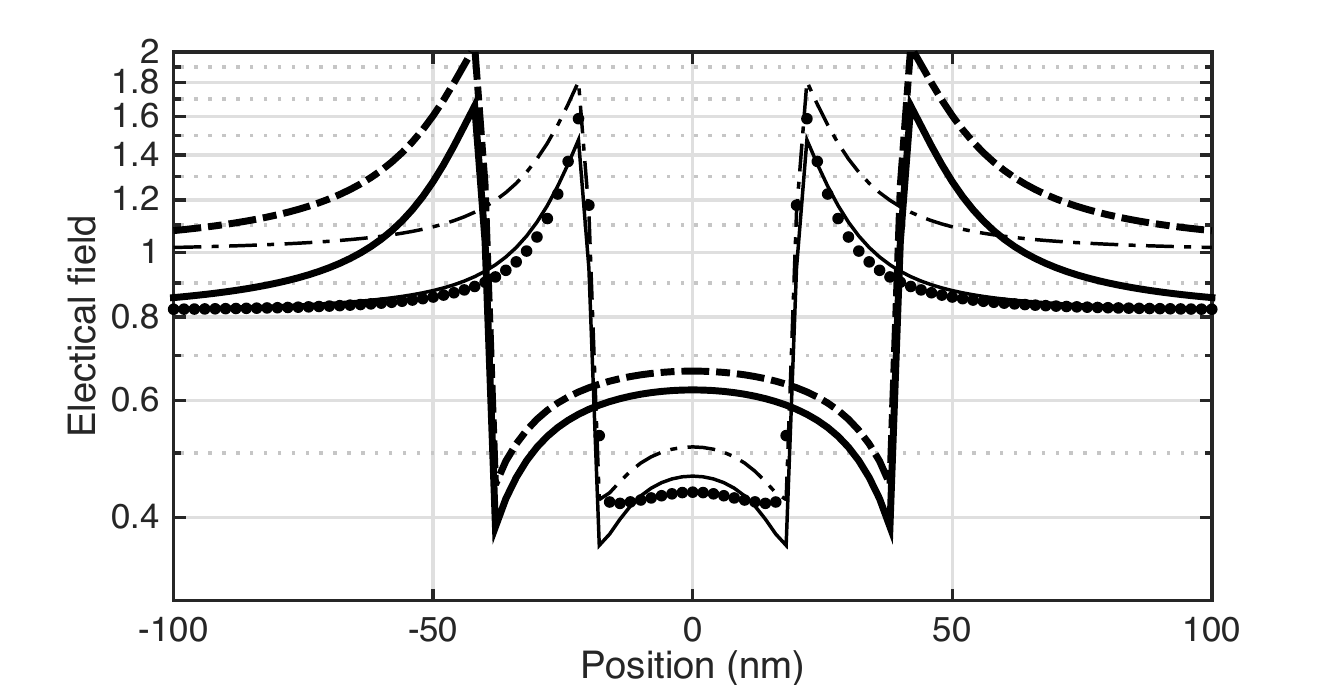} 
   \caption{Electrical field inside and outside of dielectric disks ($n=2.4$). Simulations are shown for disks on a substrate ($n=1.46$, black solid lines) and without substrate (black dashed lines). The thickness of the disks is 20 nm in all cases, while the diameters are 80 nm and 40 nm. The dotted line shows results for an ellipsoid $a=b=40\text{ nm}, c=20\text{ nm}$, with axis $c$ being perpendicular to the substrate. }
   \label{fig:disk_simul}
\end{figure}

An alternative technique of finding $\sigma$ has been proposed in \cite{pulsed_excit}. The "short-pulse method" is based on a sequence of very short excitation pulses, much shorter than all relaxation times from $^3\mathrm{E}$ to the lower electronic orbitals but  longer than the relaxation time of the vibrationally excited states (this relaxation path is shown by a fat dashed line in Fig.\ref{fig:NV_model}). If the  interval between the consecutive pulses  is much longer than the relaxation times from the higher electronic orbitals,  then the population of the ground electronic orbital equals  1 before every exciting pulse and $\rho_\mathrm{T}+\rho_\mathrm{T}^*=1$ for the duration of the pulse. In such a case, the population $\rho_\mathrm{T}^*=1-\exp(-\mathcal{E}\sigma)$ at the end of the pulse, where $\mathcal{E}$ is the number of photons carried through a unit area by one laser pulse at the location of the center.  This method has been used to determine the cross-section in a nanocrystal. The reported value of $\sigma_\mathrm{[nc]}=(9.5\pm2.5)\times10^{-17}\text{ cm}^2$ is about 3 times larger than the value cited above for a  bulk  crystal. This is a very large mismatch especially considering that the cross-section of NV-centers in nanocrystals should be smaller than in  bulk  diamond according to the above analysis (except for very flat flakes).  To explain/reduce the mismatch, the effect of the substrate on the electrical field inside a nanocrystal has been critically examined. 
\begin{figure}[htbp] 
   \centering
   \includegraphics[width=10cm]{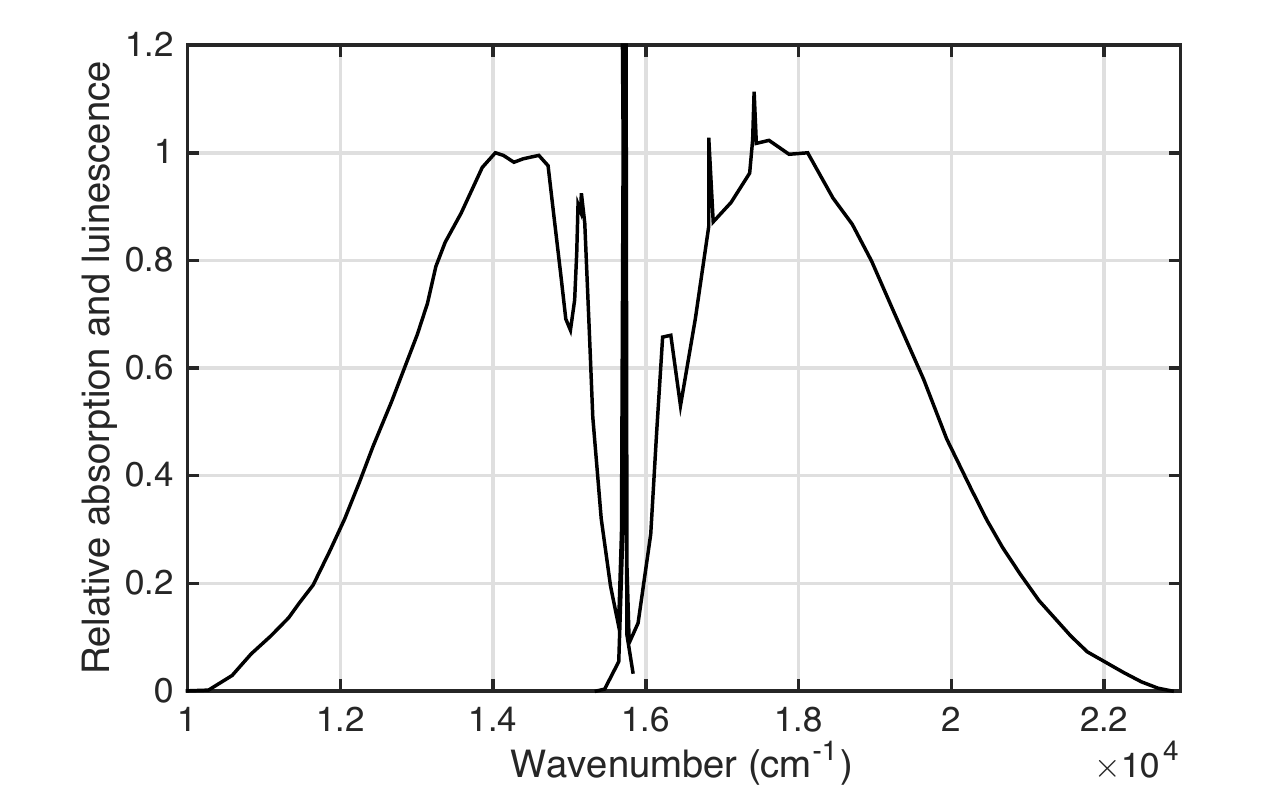} 
   \caption{Normalised (the maxima of the wide bands are set to one) of luminescence and absorption of NV-centers in diamond. A good mirror symmetry supports the validity of the Frank-Condon approximation.}
   \label{fig:lum_abs}
\end{figure}
In the absence of a nanocrystal on the surface, the reflection from the  surface interferes with the incoming wave and the resulting field is reduced in comparison to the field of the incoming wave by the factor $\eta_\mathrm{[b]}$ calculated for the material of the substrate. This reduced field has been used in \cite{pulsed_excit} as the external field for the nanocrystal (see Eq.(\ref{eq:ellipsoid_field})).  But numerical simulations (done using \emph{ Lumerical  FDTD Solutions}) quite surprisingly show that such approximation is valid only for a flake.   In less extreme cases of the aspect ratio, the substrate has a small effect on the strength of the electrical field inside the crystal. This is demonstrated in Fig.\ref{fig:disk_simul}.  Therefore the  cross-section estimated in  \cite{pulsed_excit} should be multiplied by  $\eta^2_\mathrm{[b]} = 0.66$ to eliminated the incorrectly included interface effect. 

\begin{figure}[htbp] 
   \centering
   \includegraphics[width=6 cm]{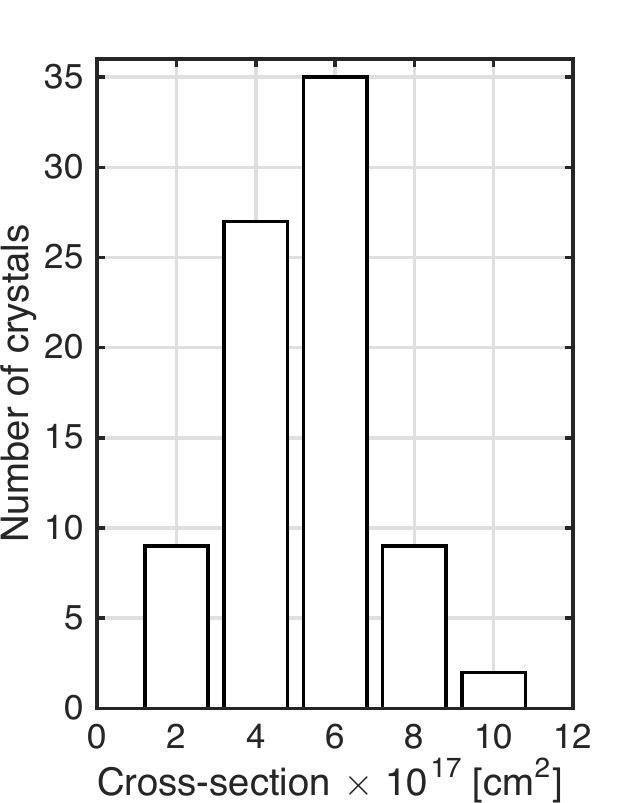} 
   \caption{Histogram of the absorption cross-section measured for 82 diamond nanocrystals (average size of about 40 nm) on a silica substrate measured with the short-pulse method. The mean value of the cross-section is $5.1\times 10^{-17}\text{ cm}^2$ and the standard deviation of the distribution is $1.8\times 10^{-17}\text{ cm}^2$. The mean value is about 1.6 times larger than the value reported for  bulk  absorption in \cite{abs_bulk}. }
  \label{fig:cross-section_hist}
\end{figure}

Second, as follows from the Tab.\ref{tab:ellipsoid}, the cross-section depends very significantly on the shape of the crystals. The HTHP nanodiamond is obtained by crashing  larger crystals which can naturally break along easy cleavage crystallographic planes \cite{cleavage}, \{1,1,1\} being the most common. A nanocrystal can then have its largest flat surface bound to the plane of the substrate (mainly due to Van der Waals and/or electrostatic forces). In such a case, one can estimate the value of the cross-section to be in the range $0.25\sigma_\mathrm{[b]} \le \sigma_\mathrm{[nc]}\le2.4 \sigma_\mathrm{[b]}$ according to the results listed in Table \ref{tab:ellipsoid} starting from a sphere and below. This variation introduces quite a large uncertainty when the cross-sections in a  bulk  and in a nanocrystal are compared. In practice, the expected variation is smaller than the range stated above. For example, we  exclude the cases of very large aspect ratios as unrealistic. But the spread of the cross-sections is still significant.  Also note that the average cross-section is larger in  bulk  for all cases except for very thin flakes. To get experimental confirmation for the variations of the cross-section, its value has been measured with the "short-pulse method" for a large number of crystals and the results are presented in Fig.\ref{fig:cross-section_hist}. These results generally agree with the value published in \cite{pulsed_excit} once it is corrected for the factor of 0.66. A large standard deviation of the measured values is in agreement with the dependence on the shape of the crystals but the average value of the distribution ($\sigma_\mathrm{[nc]}=5.1\times10^{-17}\text{ cm}^2$) is still larger than the one reported for  bulk .  

The remaining disagreement between the  bulk  measurements and nanocrystals can be explained either by assuming inaccurate calibration of the concentration of the NV centers in  bulk  or by considering dynamics of populations more sophisticated that the one assumed in  \cite{pulsed_excit}. For example, inclusion of the sig-zag path (see Fig.\ref{fig:NV_model}) changes the dependence of the population on the pulse energy to $\rho_\mathrm{T}^*=1-\exp(-\mathcal{E}(\sigma+\sigma'))$ as follows from the rate equation $ \dot\rho_\mathrm{T}^*=\sigma\rho_\mathrm{T} - \sigma' I \rho_\mathrm{T}^*$ and $\rho_\mathrm{T}=1-\rho_\mathrm{T}^*$. Thus, the measured value of the cross-section can be larger than $\sigma$. The choice between the two options is outside the scope of this paper. For the following discussion we adopt $\sigma=(3.1\pm0.8)\times 10^{-17}\text{ cm}^2$ as the  bulk  value. The corresponding values for nanocrystals  in water and on a glass substrate are  $(2.0\pm0.5)\times 10^{-17}\text{ cm}^2$ and $(1.3\pm0.3)\times 10^{-17}\text{ cm}^2$ respectively. Here we  assume the shortest axis of the ellipsoid  is not shorter than 1/2. Other cases can be analysed with a help of Tab.\,\ref{tab:ellipsoid}.

The absorption cross-section can be related to the radiative emission rate with Eq.(\ref{eq:rad_to_sigma}). Fig.\,\ref{fig:lum_abs} shows  an example of absorption and emission spectra adopted from a classical paper \cite{Davies_1976}. Substitution of $(3.1\pm0.8)\times 10^{-17}\text{ cm}^2$ in Eq.(\ref{eq:rad_to_sigma}) results in the radiative rate $k_\mathrm{r[b]}=(44\pm11) \text{ MHz}$. This value is about two times smaller than the total decay rate $k\approx 80\text{ MHz}$ frequently reported for  bulk  samples (the earliest citation for this value is probably \cite{Collins_1983}) and suggests that  $\Phi_\mathrm{NV}$, the quantum yield of luminescence  is about 50\% while a common statement about NV-centers is that $\Phi_\mathrm{NV} =  1$.  A chain of citations  \cite{NV_rev, NV_Science_1997, QY_1994} for this value of $\Phi_\mathrm{NV}$  ends up at a footnote with a reference to an unpublished work. The footnote only claims  that the intersystem  crossing from the lowest level of $^3\mathrm{E}$ to the metastable $^1\mathrm{A}$ is very slow (78 kHz) in comparison to the rate from $^3\mathrm{E}$ to $^3\mathrm{A}$. Therefore the estimate $\Phi_\mathrm{NV} \approx 1$ holds only if the relaxation from $^3\mathrm{E}$ to $^3\mathrm{A}$ is assumed to be purely radiative. Thus, the quantum yield of NV luminescence is an opened question which we will now discuss using the data collected in Tab. \ref{tab:experimental} and Eq.(\ref{eq:sigma_qy}).  

A quick glance at the table reveals that at least one of the required  parameters is frequently not reported in the literature (most frequently missing is the quantum yield of the photo detector) but the top two references  indicate that the quantum yield in  bulk  estimated with   Eq.(\ref{eq:sigma_qy}) is about 0.5. Interestingly, a more sophisticated method used in \cite{NV_silica_fiber} results in even a smaller value. 

Next, we look at nanocrystals and the results published in \cite{PMMA_film} where nanocrystals have been distributed on a silicon substrate and then tested before and after being covered with a PMMA film. The experimental results apparently point at the yield close to 1 as the decay constant $k$ and the value of $\mathcal{R}^\mathrm{(det)}$ increase by the same factor of 1.5 upon coating. But the assumption $\Phi_\mathrm{NV} = 1$ makes it hard to explain why the proximity of a medium with a very high refractive index (3.87 for silicon) does not increase the emission rate   above the value typically reported for a glass substrate ($n=1.5$). Such independence on the refractive index can only be explained if the contribution of the radiative rate to the total value of the luminescence decay rate is small. Finally, the last raw in  the table presents quantum yield values estimated  using a change in the decay rate due to proximity of a spherical mirror. In these experiments, the nanocrystals have been covered with a film of  glass and therefore the expected reduction in the radiative rate has been smaller than in the case  of vacuum/air surroundings. Nevertheless the reported in \cite{nc_in_glass} estimate of the quantum yield is smaller than 1 with good margins (even though the uncertainties are large). In the same paper, no change in the luminescence lifetime is reported when 25-nm crystals are surrounded with air and liquids of different refractive indexes, in agreement with the assumption of a relatively small contribution which the radiative rate makes to the total luminescence decay rate. 

Bearing in mind all the uncertainties and using $44\text{ MHz}$ as a reference  for the radiative rates of NV centers in  bulk ,  one can theoretically estimate these rates  in nanocrystals to be $8\pm4\text{ MHz}$ if the crystals are immersed in water and  $5\pm3\text{ MHz}$ when deposited on a glass substrate (see the discussion below Eq.\,(\ref{eq:k_rad}) and use a factor of 1.7 to convert the rate in air to the rate on glass). The conclusion about a slower than previously claimed radiative decay rates  is also supported  by papers  \cite{spin_depol, spont_emission_surf} where  luminesce decay rates  are reported to be as low as 18 MHz for some nano crystals on a glass substrate.  The radiative rates must then be even smaller.

\begin {table} 
\begin{tabular}{ l | c | c | c| c | l | c | c }
  \hline			
  $R_\infty^\mathrm{(det)} $&NA& $\Phi_\mathrm{reg}$ &$k$ &$k_\mathrm{TS}$ &$I_\mathrm{s}$&Ref. &$\Phi_\mathrm{NV}$\\
  MHz&  & &MHz	&MHz&kW\,cm$^{-2}$& \\
  \hline
 $0.71$     & fiber   	& $0.26\pm0.1$           		&  $64\pm4$	&-- 			&130 			&\cite{NV_silica_fiber}	&$0.30\pm0.1^*$  \\
 $0.016$   & 0.75     	& $(2.3\pm1)\times10^{-3}$	&			&--			&170				&                        		& $0.6\pm0.25^*$\\
    		&     		& 						&			&			&				&                        		& $0.18 \pm0.07$\\
  \hline
 $0.004$   & 0.85       & $1.5\times10^{-4}$		&  50 		&33			&360 			& \cite{stab_source}     	& $0.9^*$ \\
  \hline  
 2.4--4.6 	&bullseye  &--  						&--			&--			&--				& \cite{bullseye}            	& -- \\
 $\approx0.2$& 1.3 	&-- 						&--			&--	        		&--  				&                           		& --\\
  \hline   
 $0.032$     & 1.3       & --      					&  86           	&--			& --				&\cite{nanodia_nonclass}	&  --  \\
 $0.022$(n) & 1.3      & 						&  40     		&--       		& --				&                            		&\\
     \hline  
 $0.10$ (n)& 0.95      &--						&$40^\dagger$	&-- 			& --    			&\cite{PMMA_film}		& -- \\
 $0.15$ (n)& 0.95      &--						&$60^\ddagger$&-- 			& --   			& 					& -- \\
  \hline  
 $0.40$ (n)& 1.35    &      --  					&  --  		&--			& --   			&\cite{parabolic_mirror}	& -- \\
 $2.0$   (n)&hemisph.& --						&  --   		&--   			&  --  			& 				   	& \\
  \hline  
 $0.021$  	& 0.95   	&--						& 85  		&--			&--  				&\cite{nanowire}		& --\\
 $0.17$	&wire    	&--						&  71  		&33			&--				&					& --\\
 \hline  
 -- (n) 	&0.9		&   						&33			&--  			&				&\cite{nc_in_glass}		&$0.09-0.58^{**}$\\
  \hline  
\end{tabular}
\caption {Experimental values are measured on a single center.  
 $^*$Value estimated using Eq.(\ref{eq:sigma_qy}). $^ \dagger$Diamond nanocrystal on Si surface ($n=3.875+0.0111i$) and $^ \ddagger$when they are covered with 111-nm thick PMMA film ($n=1.479$). The rates have been measure for 10 crystals and the range of the distribution is $34\text{ MHz}<k<90\text{ MHz}$  and $50\text{ MHz}<k<130\text{ MHz}$ for coated with PMMA and uncoated crystals. (n) indicates that data are measured for a nanocrystal. $^{**}$Nanocrystals embedded in glass and the range of $\Phi_\mathrm{NV}$ is determined by modifying the radiative rate with an approaching spherical mirror.} \label{tab:experimental} 
\end{table}

So far we have mostly concentrated on radiative rates of NV-centers. In this paragraph, we briefly discuss the not radiative rates which also affect  the quantum yield and is an important factor for applications.   Comparison of luminescence of nanodiamonds suspended in water to Raman scattering of water opens an interesting way to calibrate the sensitivity of the detection system and to find the absolute intensity of luminescence \cite{Plakhotnik_2013}. It turns out \cite{NV_in_water} that Raman scattering of water is about 55 times weaker  than luminescence of 100-nm  crystals suspended at concentration of 0.1 g/L. But a unit volume of the same material starts to emit about 25 times less photons when the crystals are crashed to about 35-nm size. The reason of this reduction is not clear and only partially can be caused by destruction of the centers in the crashing process.  In part, it can be attributed to an increased significance of luminescence intermittency (also called "blinking") \cite{NV_blikning_2010} which effectively reduces the number of optically detectable centers. The onset of blinking has been observed \cite{NV_blinking} once the location of NV-centers approach  the crystals surface closer than 8 nm  and therefore blinking is more significant for smaller crystals. Another reason for decreased luminescence  is increased not radiative rates in tinier crystals. The not radiative  pathways of energy relaxation processes in nanocrystals  are enhanced in comparison to  bulk  by the proximity of the crystal surface \cite{DRM_plakhotnik_2010}. The distribution of the not radiative rates within an ensemble of   NV-centers and/or nanocrystals  (in addition to the wide distribution of the radiative rates as discussed in this paper) make any prediction of the quantum yield quite unreliable. The reported luminescence life time of NV-centers in nanocrystals on a glass substrate  is typically between 10 and 30 ns \cite{spont_emission_surf, SDN_diamond_lifetime}. Using these limits and the above estimated range for the radiative rates, one can expect the luminescence quantum yield of NV-centers in a nanocrystal on glass to be between  0.02 and 0.25. The yield would be even smaller if the crystals were completely surrounded by air.

\section{Conclusion}
Basic photo physical properties of the NV-centers in diamond revised in this paper can be summarised as follows. The commercially available nanocrystals are characterised by a very large (more than 3 orders in magnitude) spread of brightness. However, their specific brightness (the maximum number of emitted photons per unit of volume) seems to be consistent across the batch and equals approximately $1.5\times10^3$ photon/nm$^3$ for the tested samples. Thus, the main reason for the spread of brightness is the large variance of the distribution of crystal volumes.  The analysis of data presented in this paper and published elsewhere supports the conclusion that the radiative rate of NV-center luminescence decay is significantly smaller than the usually observed total rate of depopulation of the excited state. In particular, the radiative rate of NV-centers in nano diamonds (50 nm and smaller) is about 8 MHz for crystals immersed in water and around 3 MHz if the crystals are suspended in a media with refractive index close to 1 (air, vacuum). There is a very wide (deviation by a factor of 4) intrinsic  distribution of the radiative rates due to the commonly observed variety of the crystal shapes.  The radiative rate of NV-centers in  bulk  diamond is much higher and is estimated to be around 44 MHz.  These results support a revised estimate for the luminescence quantum yield which is likely to be significantly smaller than 1 for  bulk  crystals (the most probable  value is 0.5)  and is much smaller than 1 for nanocrystals.  For example, in the case of nanocrystals immersed in water, the luminescence yield of NV-centers  is typically smaller than 0.2 (but can be higher in some of the crystals with an appropriate shape and orientation of the transition dipole moment). The theoretical considerations suggest, for example, that shaping the nanocrystals in a form of flakes will improve  the yield. The absorption cross-section of the NV-centers is also smaller for the case of nanocrystals in comparison to  bulk  diamond. A larger value measured with the use of short-pulse method suggests that there can be an additional path of photo-induced depopulation of the excited triplet state via higher electronic states.  

\section{Acknowledgments}
Funding: This work was supported by the Australian Research Council [grant number  DP0771676]. 

\bibliographystyle{elsarticle-num} 

\end{document}